\documentclass[preprint,prd,aps,article,nofootinbib]{revtex4}
\usepackage{graphicx}
\usepackage{diagbox}
\usepackage{float}
\usepackage[colorlinks,
            linkcolor=black,
            anchorcolor=blue,
            citecolor=black
            ]{hyperref}
\usepackage{doi}
\usepackage{color}
\usepackage{slashed}
\begin{document}
\title{Mass spectra and wave functions of toponia}
\author{Guo-Li Wang$^{1,2,3}$, Tai-Fu Feng$^{1,2,3}$, Ya-Qian Wang$^{1,2,3}$\footnote{corresponding author}}

\affiliation{&&${^1}$ Department of Physics, Hebei University, Baoding 071002, China
\nonumber\\$^{2}$ Hebei Key Laboratory of High-precision Computation and Application of Quantum Field Theory, Baoding 071002, China
\nonumber\\$^3$ Hebei Research Center of the Basic Discipline for Computational Physics, Baoding 071002, China}

\begin{abstract}
In this article, {we solve the instantaneous Bethe-Salpeter equation with Cornell potential and Coulomb potential} and conduct a meticulous study of the mass spectrum and wave function of toponium. Our investigation reveals that, owing to the exceedingly heavy mass of the top quark, the mass splitting between singlet and triplet states, as well as within the triplet states, is negligible. Consequently, relativistic corrections can be safely disregarded in the study of toponium. As such, we present the nonrelativistic wave functions for $S$-wave, $P$-wave, and $D$-wave toponia and study the decays $\eta_t\to \gamma\gamma$, $\eta_t\to gg$, and $\Theta\to \ell^+\ell^-$.
\end{abstract}
\maketitle

\section{Introduction}
The top quark, known for its extreme mass and rapid decay, exhibits a broad width of $1.42^{+0.19}_{-0.15}$ GeV~\cite{pdg}. Consequently, it has been widely accepted that there is insufficient time for the top quark to form a bound state before undergoing decay. As a result, the investigation into the properties of toponium bound states has received limited attention, with only a handful of studies addressing this topic. {For instance, some research has explored the production of top quarkonium \cite{Pancheri}, the impact of bound state effects on toponium near threshold in top-quark pair production at the LHC ~\cite{Sumino,Kawabata,WanliJu,Fabio} and future ILC ~\cite{Fadin,ma}, a $1S$ peak below the threshold of top-quark pairs is reported~\cite{Kaoru,Kiyo}, as well as delved into the mass spectrum~\cite{Buchmuller,Moxhay,penin,Kiyo2} or the binding energy \cite{Fadin}, wave function or wave function at the origin of $S$-wave toponium ~\cite{beneke,Fabiano2}, two-gamma decay or two-gluon decay of pseudoscalar $\eta_t$~\cite{Pancheri,Fabiano2,Kuhn1,Kuhn2,Fabiano,Fabiano1,Cakir,Kats}, and dilepton decay of the vector $\Theta$~\cite{Buchmuller,Bigi,Artymowicz,Moxhay,Yndurain} are also studied with different masses of toponium, etc. Interested readers can refer to Ref. \cite{Kuhn1}, which provided a comprehensive introduction to the various aspects of toponium.}

{
Recently, there have been significant developments in this area. The ATLAS  Collaboration has reported the highest-energy observation of quantum entanglement in top quark pairs ~\cite{atlas}. The measured entanglement marker is $D=-0.547\pm0.002$(stat.)$\pm0.021$(syst.) for $340<m_{t\bar t}<380$ GeV. The observed result is more than five standard deviations from a scenario without entanglement. The CMS  Collaboration confirmed the quantum entanglement in top quark pairs ~\cite{cms}. These experimental discoveries have stimulated theoretical interest in the studying of quantum entanglement of $t\bar t$ pairs \cite{han,Dong,Casas,han2}.

In the observation of the entanglement in top quark pairs, the ATLAS Collaboration has also identified a slight deviation between their data and the Standard Model prediction. A few year ago, both ATLAS Collaboration and CMS Collaboration found difficulties in modeling the data near the $t\bar t$ production threshold \cite{atlas2,cms2}, indicating including toponium in $t\bar t$ production simulations is important and critical. In Ref.~\cite{cms}, the CMS Collaboration showed that their results align well when considering the contribution of the pseudoscalar $t\bar t$ bound state, $\eta_t$, in the signal model. This suggests a potential hint of the existence of $\eta_t$. Furthermore, there have been theoretical studies on the possible discovery and characterization of toponium production at the LHC ~\cite{ma2,Aguilar}, indicating a growing interest in this area of research \cite{Estrada,Francener}. More recently, the CMS Collaboration has initiated a search for heavy scalar or pseudoscalar spin-0 states in $t\bar t$ events  with 138 ${\rm fb}^{-1}$ proton-proton collision data at $\sqrt{s}=13$ TeV. An excess is observed in the low bins of $m_{t\bar t}$, and is found to be consistent with the pseudoscalar hypothesis \cite{2411}.

Given this evolving landscape, and considering that the existing literature on toponium (prior to the discovery of the top quark) predominantly assumed a top quark mass far below the experimentally determined value of 172.7 GeV, it is crucial to carefully investigate the fundamental properties of toponium with top quark mass of $m_t=172.7$ GeV, such as its mass spectrum and wave function. Therefore, in this article, we solve the instantaneous Bethe-Salpeter (BS) equation ~\cite{Salpeter:1951sz}, also known as the Salpeter equation ~\cite{Salpeter:1952ib}, to study the mass spectrum of toponium and to provide the corresponding wave functions. These derived wave functions are further employed to examine the decays of the pseudoscalar $\eta_t$ into two photons and two gluons, as well as the vector $\Theta$ to dileptons.}

The article is organized as follows, the Sec.II presents the mass spectra and wave functions, Sec.III calculates the decays of $\eta_t\to \gamma\gamma$, $\eta_t\to gg$, and $\Theta\to \ell^+\ell^-$, and Sec.IV offers a brief discussion.

\section{Mass spectra and wave functions}
{To calculate the mass spectra and wave functions of $t\bar t$ bound states, we solve the instantaneous BS equation, i.e., the Salpeter equation. Since we have already solved the Salpeter equation for mesons with different $J^P$ in previous articles, we will not provide a detailed introduction to the equation-solving process here. For readers' reference, we have included brief descriptions of both the BS equation and the Salpeter equation in the Appendix. Those interested in the equation-solving methodology may consult Refs. \cite{Kim:2003ny,Wang:2005qx} for details. In this paper, when solving the Salpeter equation we first consider the linear potential plus Coulomb potential, which is also called Cornell potential \cite{cornell}}, given by
\begin{equation}
V_{\rm Cornell}(r)=\lambda r-\frac{4}{3}\frac{\alpha_s}{r},\label{potential}
\end{equation}
where $\lambda=0.18$ GeV$^2$ represents the string tension, and $\alpha_s$ denotes  the
running strong coupling constant. In momentum space, incorporating one-loop QCD correction, we express $\alpha_s(\vec{q})$ as~{\cite{Richardson,ktchao}}
$$\alpha_s(\vec{q})=\frac{12\pi}{33-2N_f}\frac{1}
{\log (e+\frac{{\vec{q}}^2}{\Lambda^{2}})},$$
where $e=2.71828$, $N_f=5$ for the $t \bar t$ system, and $\Lambda=0.10$ GeV. Utilizing these parameters along with the constitute quark mass $m_t=172.7$ GeV, we determine $\alpha_s(m_t)=0.11$. {Secondly, as a comparison to the Cornell potential, and it was pointed out that the short distance Coulomb potential plays a dominant role in the system of toponium \cite{Fadin,Fabiano1,ma} since the radius of toponium is very small \cite{Goncharov}, so we also calculate the case with only Coulomb potential and ignore the linear one, that is,
\begin{equation}
V_{\rm Coulomb}(r)=-\frac{4}{3}\frac{\alpha_s}{r}.
\end{equation}}

With the specified interaction potential and relativistic wave function forms as input, we solve the full Salpeter equations. For example, the input form of the pseudoscalar wave function is given by~\cite{Kim:2003ny}
\begin{equation}\label{e_0-}
\Psi_{_P}^{0^{-}}({q}_{_\bot})=\displaystyle
\left(\phi^{}_1 M+\phi^{}_2\not\!P+\phi_3\not\!{q}_{_\bot}
+\phi_4\frac{\not\!{q}_{_{\bot}}\not\!P}{M}\right)\gamma^{5},
\end{equation}
where $P$ and $M$ are the momentum and mass of the toponium, respectively; $q$ is the relative momentum between the quark momentum
$p_{_1}=\frac{1}{2}P+q_{_{\bot}}$ and antiquark momentum  $p_{_2}=\frac{1}{2}P-q_{_{\bot}}$, with ${q}_{_{\bot}}=q-\frac{P\cdot q}{M^2}P$ defined accordingly.
In the center of the toponium mass system, $q=(0,\vec{q})$, where $\vec{q}$ is the momentum vector. The radial part of the wave function, denoted by $\phi_i$ for $i=1,2,3,4$, is a function of $-{q}_{_{\bot}}^2=\vec{q}^2$, and its numerical value is determined by solving the full Salpeter equation.

The vector wave function is expressed as per Ref.~\cite{Wang:2005qx}
$$\Psi_{_P}^{1^{-}}({q}_{_{\bot}})=
{\epsilon}\cdot{q}_{_{\bot}}
\left[\psi_1+\frac{\not\!P}{M}~\psi_2+
\frac{{\not\!{q}_{_{\bot}}}}{M}~\psi_3+\frac{{\not\!P}
{\not\!{q}_{_{\bot}}}}{M^2}~\psi_4\right]+
M{\not\!\epsilon}~\psi_5$$
\begin{equation}+
{{\not\!P}\not\!\epsilon}~\psi_6+
({\not\!{q}_{_{\bot}}}{\not\!\epsilon}-
{\epsilon}\cdot{q}_{_{\bot}})
~\psi_7+\frac{1}{M}({\not\!P}{\not\!\epsilon}
{\not\!{q}_{_{\bot}}}-{\not\!P}{\epsilon}\cdot{q}_{_{\bot}})
~\psi_8,
\label{e_1-}
\end{equation}
where $\epsilon$ is the polarization vector of the meson, and the radial wave function $\psi_i=\psi_i(-{q}^2_{_{\bot}})$.

We solve the full Salpeter equations \cite{Salpeter:1951sz,Salpeter:1952ib} for mesons with various $J^P$ quantum numbers individually ~\cite{Wang:2022cxy}, and obtain the mass spectra and relativistic wave functions of the toponium.
Our findings indicate that the relativistic corrections can be neglected, aligning with the substantial mass of the top quark. Additionally, we observe that the mass splitting between the singlet and triplet states is entirely negligible due to the heavy top quark mass. For instance, we find
\begin{equation}
M_{0^{-+}}(n^1S_0)= M_{1^{--}}(n^3S_1),
\end{equation}
where $n$ is the principle quantum number. The notation $J^{PC}$ and $n^{2S+1}L_J$ are utilized to differentiate between various bound states.
Moreover, the heavy top quark leads to a negligible mass splitting within the triplet states, and we have
\begin{equation}
M_{0^{++}}(n^3P_0)= M_{1^{++}}(n^3P_1)= M_{2^{++}}(n^3P_2)= M_{1^{+-}}(n^1P_1),
\end{equation}
\begin{equation}
M_{1^{--}}(n^3D_1)= M_{2^{--}}(n^3D_2)= M_{3^{--}}(n^3D_3)= M_{2^{-+}}(n^1D_2),
\end{equation}
\begin{equation}
M_{2^{++}}(n^3F_2)= M_{3^{++}}(n^3F_3)= M_{4^{++}}(n^3F_4)= M_{3^{+-}}(n^1F_3),
\end{equation}
and
\begin{equation}
M_{3^{--}}(n^3G_3)= M_{4^{--}}(n^3G_4)= M_{5^{--}}(n^3G_5)= M_{4^{-+}}(n^1G_4).
\end{equation}

{The calculated masses of toponia with Cornell potential and Coulomb potential are shown in Table~\ref{tab1} and Table~\ref{tab2}, respectively, with states having the same mass indicated within parentheses. A comparison of the two tables reveals that the masses predicted by the Cornell potential are slightly larger than the corresponding values from the Coulomb potential, indicating that the Coulomb potential binds heavy quarkonium systems more deeply. Additionally, we note that the mass splitting obtained using the Cornell potential is also slightly larger than that derived from the Coulomb potential.}
\squeezetable
\begin{table}[hbt]
\caption{Mass spectra of toponia $t\bar t$ in unit of GeV with Cornell potential.} \label{tab1}
\begin{tabular}{cccccccccc}
 \hline\hline
~~$n^{2S+1}L_J$~~&~~M~~&~~$n^{2S+1}L_J$~~&~~M~~&~~$n^{2S+1}L_J$~~&~~M~~&~~$n^{2S+1}L_J$~~&~~M~~&~~$n^{2S+1}L_J$~~&~~M~~\\ \hline
$1^{1}S_0$(${}^{3}S_1$)~~&~~343.62~~&~~$1^{1}P_1$(${}^{3}P_{0,1,2}$)~~&~~344.39~~&~~$1^{1}D_2$(${}^{3}D_{1,2,3}$)~~&~~344.72~~&~~$1^{1}F_3$(${}^{3}F_{2,3,4}$)~~&~~344.93~~&~~$1^{1}G_4$(${}^{3}G_{3,4,5}$)~~&~~345.07\\
$2^{1}S_0$(${}^{3}S_1$)~~&~~344.59~~&~~$2^{1}P_1$(${}^{3}P_{0,1,2}$)~~&~~344.83~~&~~$2^{1}D_2$(${}^{3}D_{1,2,3}$)~~&~~344.99~~&~~$2^{1}F_3$(${}^{3}F_{2,3,4}$)~~&~~345.12~~&~~$2^{1}G_4$(${}^{3}G_{3,4,5}$)~~&~~345.22\\
$3^{1}S_0$(${}^{3}S_1$)~~&~~344.93~~&~~$3^{1}P_1$(${}^{3}P_{0,1,2}$)~~&~~345.06~~&~~$3^{1}D_2$(${}^{3}D_{1,2,3}$)~~&~~345.16~~&~~$3^{1}F_3$(${}^{3}F_{2,3,4}$)~~&~~345.25~~&~~$3^{1}G_4$(${}^{3}G_{3,4,5}$)~~&~~345.33\\
 \hline\hline
\end{tabular}
\end{table}
\squeezetable
\begin{table}[hbt]
{\caption{Mass spectra of toponia $t\bar t$ in unit of GeV with Coulomb potential.}} \label{tab2}
\begin{tabular}{cccccccccc}
 \hline\hline
~~$n^{2S+1}L_J$~~&~~M~~&~~$n^{2S+1}L_J$~~&~~M~~&~~$n^{2S+1}L_J$~~&~~M~~&~~$n^{2S+1}L_J$~~&~~M~~&~~$n^{2S+1}L_J$~~&~~M~~\\ \hline
$1^{1}S_0$(${}^{3}S_1$)~~&~~343.59~~&~~$1^{1}P_1$(${}^{3}P_{0,1,2}$)~~&~~344.33~~&~~$1^{1}D_2$(${}^{3}D_{1,2,3}$)~~&~~344.65~~&~~$1^{1}F_3$(${}^{3}F_{2,3,4}$)~~&~~344.83~~&~~$1^{1}G_4$(${}^{3}G_{3,4,5}$)~~&~~344.94\\
$2^{1}S_0$(${}^{3}S_1$)~~&~~344.52~~&~~$2^{1}P_1$(${}^{3}P_{0,1,2}$)~~&~~344.73~~&~~$2^{1}D_2$(${}^{3}D_{1,2,3}$)~~&~~344.86~~&~~$2^{1}F_3$(${}^{3}F_{2,3,4}$)~~&~~344.96~~&~~$2^{1}G_4$(${}^{3}G_{3,4,5}$)~~&~~345.02\\
$3^{1}S_0$(${}^{3}S_1$)~~&~~344.80~~&~~$3^{1}P_1$(${}^{3}P_{0,1,2}$)~~&~~344.90~~&~~$3^{1}D_2$(${}^{3}D_{1,2,3}$)~~&~~344.98~~&~~$3^{1}F_3$(${}^{3}F_{2,3,4}$)~~&~~345.04~~&~~$3^{1}G_4$(${}^{3}G_{3,4,5}$)~~&~~345.08\\
 \hline\hline
\end{tabular}
\end{table}

{Our mass predictions, whether using the Cornell potential or the Coulomb potential, show that only the $1S$ toponium has a mass below 344 GeV. This result aligns with the outcomes in Refs.~\cite{Kiyo,Kaoru}, whose calculations on the top-quark pair production indicate that the $1S$-singlet toponium contributes a peak in the $t\bar t$ invariant mass spectrum around $343\sim344$ GeV. A similar conclusion is also reported in Ref. \cite{WanliJu}. The masses listed in Tables \ref{tab1} and \ref{tab2} are all below the value of $2m_t$. Therefore, our results also show that there exist multiple possible bound states below the $2m_t$ threshold.}

In our relativistic calculations for toponium, the relativistic corrections and spin-dependent potentials are found to be negligible. Consequently, for simplicity, we will provide the expressions of the nonrelativistic wave functions for $S$-wave, $P$-wave, and $D$-wave toponia. The pseudoscalar and vector relativistic wave functions in Eq.~\ref{e_0-} and Eq.~\ref{e_1-} reduce to their nonrelativistic forms for the $S$ waves
\begin{equation}
\Psi_{_{^1S_0}}^{0^{-+}}(\vec{q})=(\slashed P+M)\gamma_{5} \phi_{_S}(\vec{q}),
\end{equation}
\begin{equation}
\Psi_{_{^3S_1}}^{1^{--}}(\vec{q})=(\slashed P+M)\slashed{\epsilon} \phi_{_S}(\vec{q}),
\end{equation}
where $\phi_{_S}(\vec{q})$ is the radial wave function.
And the normalization condition is
\begin{equation}\int\frac{d{\vec{q}}}{(2\pi)^3}
\frac{4m_tM\phi^2_{_S}({\vec{q}})}{\omega}
=1,
\end{equation}
where $\omega=\sqrt{m_t^{2}+{\vec q}^{2}}$.

There are four $P$-wave states that share the same radial wave function $\phi_{_P}(\vec{q})$ \cite{wglP,wglP2}:
\begin{equation}
\Psi_{_{^3P_0}}^{0^{++}}(\vec{q})=(\slashed P+M)\frac{{\not\!{q}_{_{\bot}}}}{M} \phi_{_P}(\vec{q}),
\end{equation}
\begin{equation}
\Psi_{_{^3P_1}}^{1^{++}}(\vec{q})=(\slashed P+M)\frac{i\varepsilon_{\mu\nu\alpha\beta}\gamma^{\mu}P^{\nu}{q}^{\alpha}_{_{\bot}}\epsilon^{\beta}}{M^2} \sqrt{\frac{2}{3}}\phi_{_P}(\vec{q}),
\end{equation}
\begin{equation}
\Psi_{_{^3P_2}}^{2^{++}}(\vec{q})=(\slashed P+M)\frac{\epsilon_{\mu\nu}\gamma^{\mu}{q}^{\nu}_{_{\bot}}}{M} \sqrt{\frac{1}{3}}\phi_{_P}(\vec{q}),
\end{equation}
and
\begin{equation}
\Psi_{_{^1P_1}}^{1^{+-}}(\vec{q})=(\slashed P+M)\gamma_{5}\frac{{\epsilon}\cdot{q}_{_{\bot}}}{M} \sqrt{\frac{1}{3}}\phi_{_P}(\vec{q}).
\end{equation}
The normalization condition is
\begin{equation}\int\frac{d{\vec{q}}}{(2\pi)^3}
\frac{4m\vec{q}^2\phi^2_{_P}({\vec{q}})}{M\omega}
=1.
\end{equation}
Similarly, the four $D$-wave states also share the same radial wave function $\phi_{_D}(\vec{q})$, and their wave functions are represented as \cite{tianh,tianh2}
\begin{equation}
\Psi_{_{^3D_1}}^{1^{--}}(\vec{q})=(\slashed P+M)\frac{{\not\!{q}_{_{\bot}}}{\epsilon}\cdot{q}_{_{\bot}}}{M^2} \phi_{_D}(\vec{q}),
\end{equation}
\begin{equation}
\Psi_{_{^3D_2}}^{2^{--}}(\vec{q})=(\slashed P+M)\frac{i\varepsilon_{\mu\nu\alpha\beta}\gamma^{\mu}P^{\nu}{q}^{\alpha}_{_{\bot}}\epsilon^{\beta\delta}{q}_{_{\bot{\delta}}}}{M^3} \sqrt{\frac{3}{5}}\phi_{_D}(\vec{q}),
\end{equation}
\begin{equation}
\Psi_{_{^3D_3}}^{3^{--}}(\vec{q})=(\slashed P+M)\frac{\epsilon_{\mu\nu\alpha}\gamma^{\mu}{q}^{\nu}_{_{\bot}}{q}^{\alpha}_{_{\bot}}}{M^2} \sqrt{\frac{2}{5}}\phi_{_D}(\vec{q}),
\end{equation}
and
\begin{equation}
\Psi_{_{^1D_2}}^{2^{-+}}(\vec{q})=(\slashed P+M)\gamma_{5}\frac{\epsilon_{\mu\nu}{q}^{\mu}_{_{\bot}}{q}^{\nu}_{_{\bot}}}{M^2} \sqrt{\frac{2}{5}}\phi_{_D}(\vec{q}),
\end{equation}
with the normalization condition
\begin{equation}\int\frac{d{\vec{q}}}{(2\pi)^3}
\frac{4m\vec{q}^4\phi^2_{_D}({\vec{q}})}{3M^3\omega}
=1.
\end{equation}

\begin{figure}[htbp]
    \centering
    \includegraphics[scale=0.3]{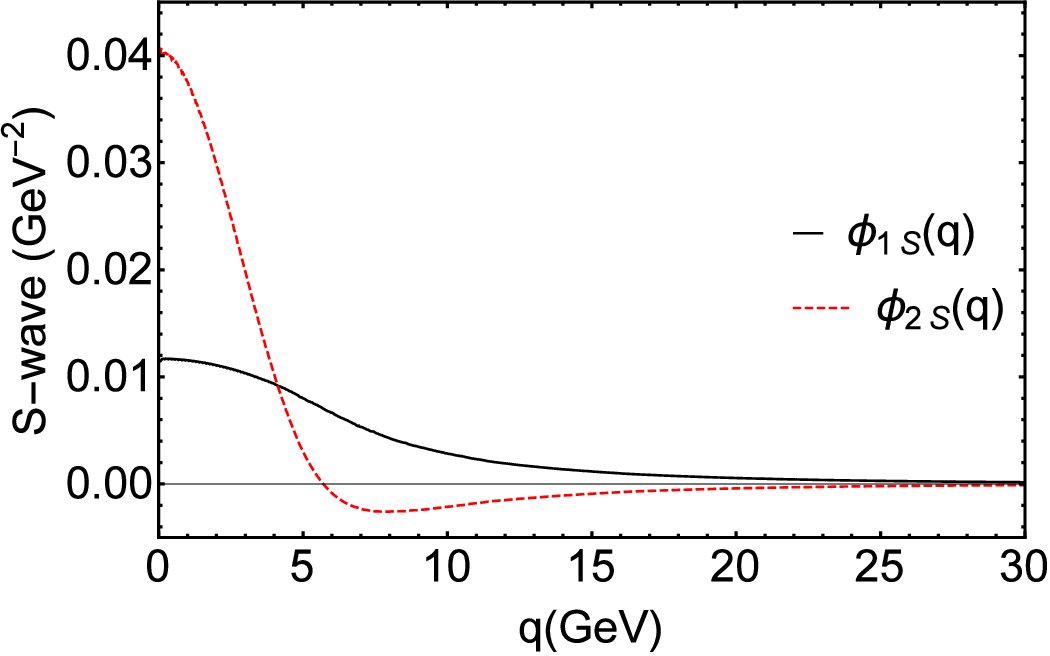}
    \includegraphics[scale=0.3]{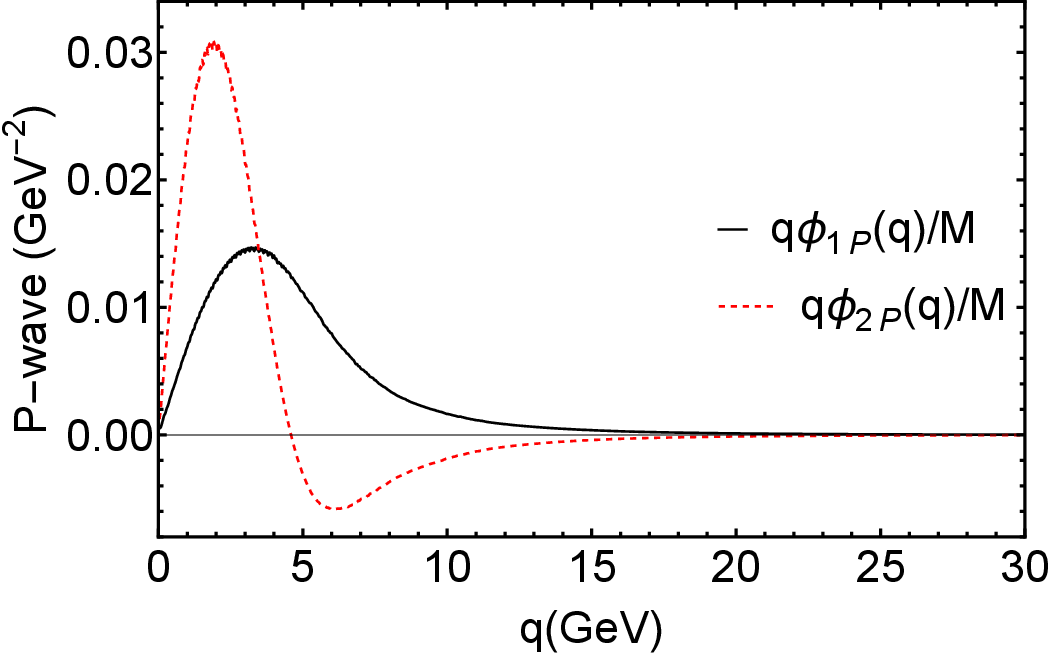}
    \includegraphics[scale=0.3]{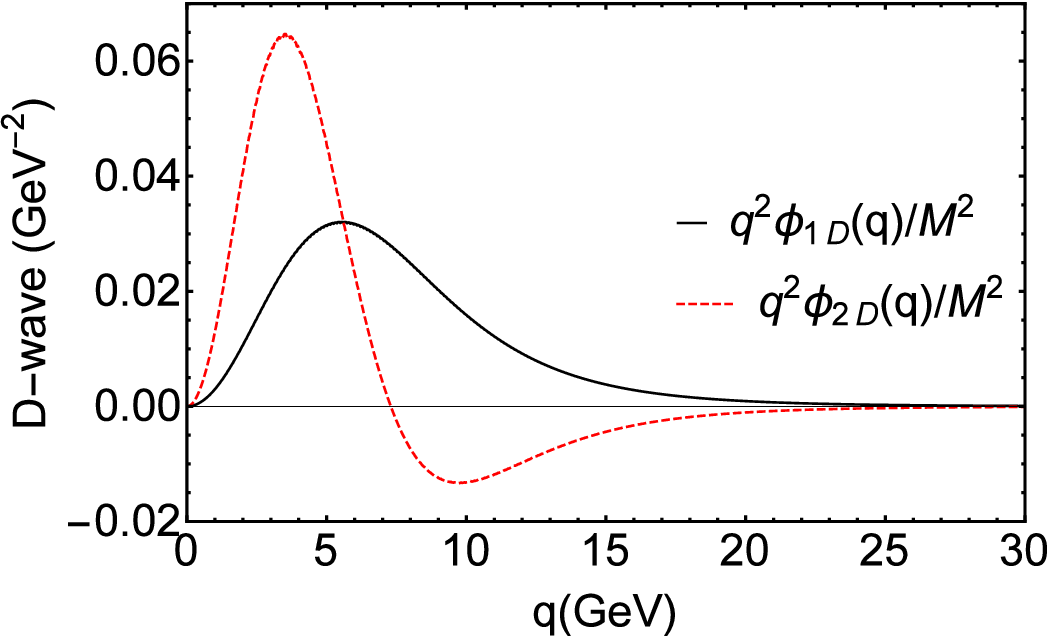}
    \caption{The radial wave functions of the ground state and the first excited state of the $S$-wave, $P$-wave and $D$-wave toponia with Cornell potential.}\label{pic1}
    \end{figure}

\begin{figure}[htbp]
    \centering
    \includegraphics[scale=0.3]{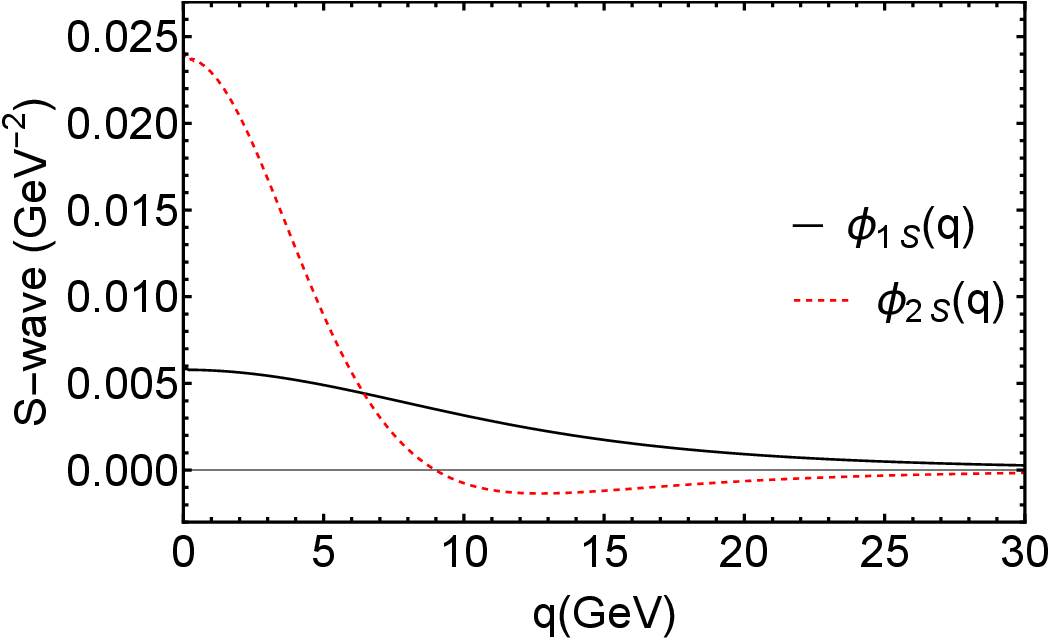}
    \includegraphics[scale=0.3]{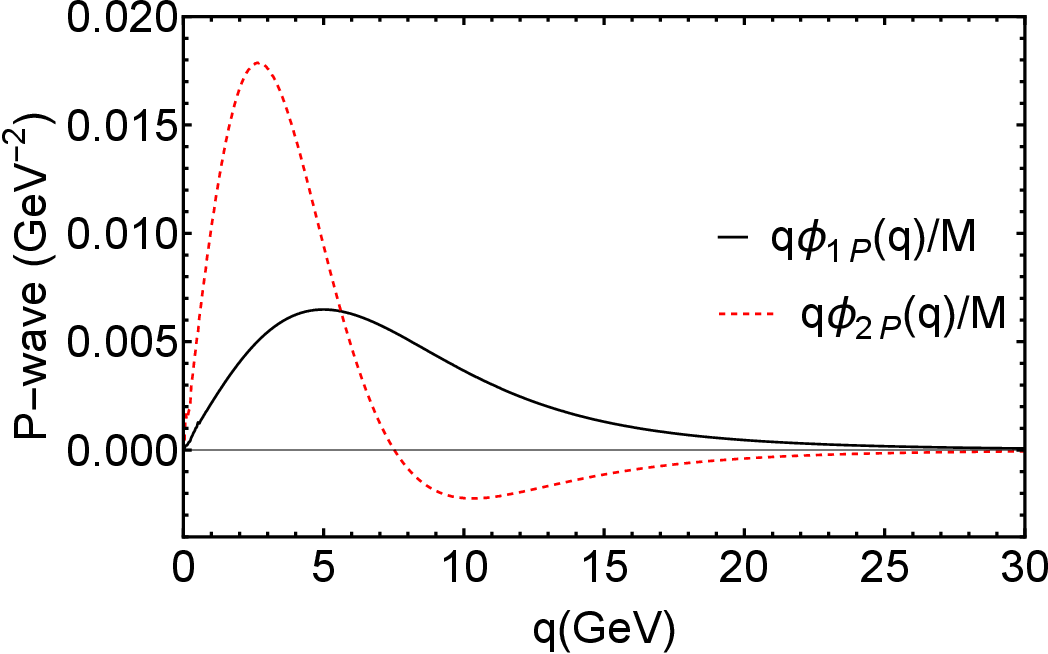}
    \includegraphics[scale=0.3]{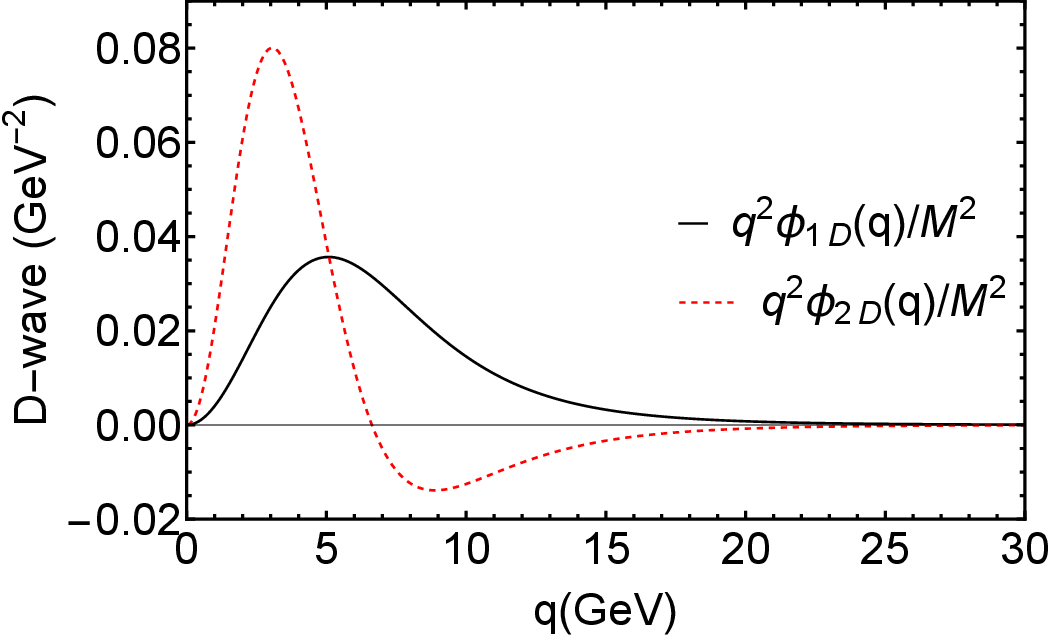}
    \caption{{The radial wave functions of the ground state and the first excited state of the $S$-wave, $P$-wave and $D$-wave toponia with Coulomb potential.}}\label{pic2}
    \end{figure}

{We plot the radial wave functions of the ground state and the first excited state for these $S$-wave, $P$-wave, and $D$-wave toponia in Fig.~\ref{pic1} using Cornell potential and in Fig.~\ref{pic2} with Coulomb potential, where $q=|\vec q|$. The results shown in the figures demonstrate that the numerical values of the radial wave functions obtained with the two different potentials exhibit significant differences. This leads to markedly divergent outcomes when applied to calculations of specific physical processes, this conclusion explicitly confirmed by the computational results presented in the following section.}

\section{Decays $\eta_t\to \gamma\gamma$, $\eta_t\to gg$, and $\Theta\to \ell^+\ell^-$}

The transition amplitude of the two-photon decay $\eta_t \to \gamma\gamma$
is given by:
\begin{eqnarray}
T & = & \sqrt{3}\; (e_qe)^2 \!
 \int \!\! \frac{d\vec{q}}{(2\pi)^3}
              \; \mbox{tr}\; \Bigg\{ \,
    \Psi^{0^{-+}}(\vec q) \bigg[\varepsilon\!\!\! /_{_2}\, \frac{1}{\not\! p_{_1}-\not\! k_{_1}}\,
    \varepsilon\!\!\! /_{_1} +
     \varepsilon\!\!\! /_{_1}\, \frac{1}{\not\! p_{_1}-\not\! k_{_2}}\,
    \varepsilon\!\!\! /_{_2} \bigg]
        \Bigg\},
\label{23}
\end{eqnarray}
where $k_{_1}$, $k_{_2}$; $\varepsilon_{_1}$, $\varepsilon_{_2}$ are the
momenta and polarization vectors of the photons, respectively; $p_{_1}$ and $e_qe=\frac{2}{3}e$ are the momentum and the charge of the top quark, respectively. The corresponding two-photon decay width is
   \begin{eqnarray}
   \Gamma(\eta_t \rightarrow \gamma\gamma)=12\pi \alpha^2
   e_q^4 M^3 \left\{
    \int \frac{d\vec{q}}{(2\pi)^3} \phi_{_S} (\vec{q})
 \left[
 \frac{1}{(p_{_1}-k_{_1})^2}+\frac{1}{(p_{_1}-k_{_2})^2} \right]
   \right\}^2~,
   \label{2gamma}
   \end{eqnarray}
where {$ \alpha=\frac{e^2}{4\pi}=\frac{1}{128}$}.

{In our calculations, we use both the Cornell potential and the Coulomb potential, and the results are presented in Table \ref{tab3}. The results show that there is a significant difference between the two, especially in the ground states. When the Cornell potential is chosen, the decay width is
\begin{equation}
\Gamma_{\rm Cornell}(\eta_t \rightarrow \gamma\gamma)=7.56~~{\rm keV},
\end{equation}
while with Coulomb potential, we obtain
\begin{equation}
\Gamma_{\rm Coulomb}(\eta_t \rightarrow \gamma\gamma)=15.9~~{\rm keV}.
\end{equation}
{Since the parameters of the Coulomb term are identical in both potentials, and the difference lies in the additional linear term in the Cornell potential, it is evident that the linear potential significantly impacts the wave function of the toponium (see Figs 1 and 2), reducing its width by nearly half.} Our result of 15.9 keV obtained using the Coulomb potential { is very consistent with Ref. \cite{Goncharov}, where a relatively crude method yielded an estimated value of 16.865 keV, and is slightly smaller than the value of 22.6 keV obtained using the nonrelativistic QCD in Ref. \cite{Enterria}. In Ref. \cite{Enterria}, the diphoton width depends on the coordinate-space wave function at origin, which is directly related to the strong coupling constant $\alpha_s$ at a fixed scale, while ours employs a running strong coupling constant, and this difference may constitute the primary reason for the discrepancy in the widths.}

\squeezetable
\begin{table}[hbt]
{\caption{Two-gamma decay width (keV), two-gluon decay width (MeV), and dilepton decay width (keV) of toponia using Cornell and Coulomb potentials.}} \label{tab3}
\begin{tabular}{ccccccccc}
 \hline\hline
channel~(keV)~~&~~Cornell~~&~~{Coulomb}~~&~~channel~(MeV)~~&~~Cornell~~&~~{Coulomb}~~&~~channel~(keV)~~&~~Cornell~~&~~{Coulomb}~~\\ \hline
$\eta_t \rightarrow \gamma\gamma$~~&~~7.56~~&~~15.9~~&~~$\eta_t \rightarrow gg$~~&~~1.69~~&~~3.54~~&~~$\Theta \rightarrow \ell^+\ell^-$~~&~~6.09~~&~~12.9\\
$\eta'_t \rightarrow \gamma\gamma$~~&~~2.69~~&~~3.13~~&~~$\eta'_t \rightarrow gg$~~&~~0.600~~&~~0.697~~&~~$\Theta' \rightarrow \ell^+\ell^-$~~&~~2.14~~&~~2.49\\
$\eta''_t \rightarrow \gamma\gamma$~~&~~1.29~~&~~1.26~~&~~$\eta''_t \rightarrow gg$~~&~~0.288~~&~~0.281~~&~~$\Theta'' \rightarrow \ell^+\ell^-$~~&~~1.03~~&~~1.01\\
 \hline\hline
\end{tabular}
\end{table}

The two-gluon decay width of $\eta_t \rightarrow gg$ can be obtained with the replacement
${e_q}^4{\alpha}^2\to
\frac{2}{9}{\alpha_s}^2$ in Eq.~\ref{2gamma},
So we have
 \begin{eqnarray}
   \Gamma(\eta_t \rightarrow gg)=\frac{8}{3}\pi \alpha_s^2
  M^3 \left\{
    \int \frac{d\vec{q}}{(2\pi)^3} \phi_{_S} (\vec{q})
 \left[
 \frac{1}{(p_{_1}-k_{_1})^2}+\frac{1}{(p_{_1}-k_{_2})^2} \right]
   \right\}^2~,
   \end{eqnarray}
{where the $\alpha_s=\alpha_s(M_{\eta_t})=0.11$. The obtained results with Cornell potential and Coulomb potential are shown in Table \ref{tab3}. Similar to the results of the two-photon decay, the results obtained with the two potentials exhibit discrepancies, particularly a significant difference in the ground-state outcomes. The Cornell potential yields
\begin{equation}
\Gamma_{\rm Cornell}(\eta_t \rightarrow gg)=1.69~~{\rm MeV},
\end{equation}
while the Coulomb potential produces
\begin{equation}
\Gamma_{\rm Coulomb}(\eta_t \rightarrow gg)=3.54~~{\rm MeV},
\end{equation}
more than twice the result obtained from the Cornell potential.
For the two-gluon decay of pseudoscalar $\eta_t$, Ref. \cite{Kats} utilized the formula from Ref. \cite{Kuhn1} and obtained a result of 4 MeV using the Coulomb potential, which is in good agreement with our result of 3.54 MeV.}

The dilepton decay width of the vector $1^{--}$ toponium $\Theta$ is given by
\begin{equation}
\Gamma_{\Theta\to
\ell^{+}\ell^{-}}=\frac{4\pi\alpha^2e_q^2F^2_V}{3M},
\end{equation}
where $F_V$ is the decay constant of $\Theta$ defined as
\begin{equation}
<0|\bar t\gamma_{\mu} t|\Theta(P,\epsilon)>=MF_V\epsilon_{\mu},
\end{equation}
with $F_{V} = 4\sqrt{3} \int \frac{d^3 \vec{q}}{(2\pi)^3} \phi_{_S}({\vec q})$.
{We also show the results of $\Theta\to\ell^{+}\ell^{-}$ with the Cornell potential and the Coulomb potential in Table \ref{tab3}. Where we can see, similar to the two-photon decay of the color-singlet ground state toponium, the dileton decay of the color-triplet ground state toponium shows significantly different results when calculated with the Cornell potential compared to the Coulomb potential. The result obtained using the Cornell potential is
\begin{equation}
\Gamma(\Theta \rightarrow \ell^+\ell^-)=6.09~~{\rm keV},
\end{equation}
less than half of the result derived from the Coulomb potential
\begin{equation}
\Gamma(\Theta \rightarrow \ell^+\ell^-)=12.9~~{\rm keV}.
\end{equation}
However, for the first and second excited states under both interaction potentials, the results exhibit little difference due to the cancellation of contributions from regions before and after the nodes in the wave functions. For the results calculated using the Coulomb potential, Ref. \cite{Yndurain} reports a two-electron partial width $\Gamma(\Theta \rightarrow \ell^+\ell^-)=13\pm 1~~{\rm keV}$, which shows excellent agreement with our results.}

\section{Discussions}
Because the top quark is extremely heavy, the behavior of bound state of $t\bar t$, toponium, is completely different from that of charmonium and bottomonium, which are also heavy quarkonia. Firstly, in toponium spectroscopy, the mass splitting between the singlet and the triplet is zero, for example, $M(n^1S_0)= M(n^3S_1)$ and $M(n^3P_J)= M(n^1P_1)$, and the mass splittings within the triplet states are also zero, such as $M(n^3P_0)= M(n^3P_1)= M(n^3P_2)$. Secondly, in terms of decays, while the two-photon decay and two-gluon decay are crucial for the ground pseudoscalar charmonium and bottomonium, with the latter having a width equivalent to the total width of the quarkonium, for pseudoscalar toponium, the partial widths of two-photon decay and two-gluon decay are relatively small compared to the total width. Similarly, the dilepton decay width of vector toponium, which plays a significant role in charmonium and bottomonium, is also very small. Conversely, weak decay contributes minimally and can be disregarded in charmonium and bottomonium decays, but it is vital and plays a dominant role in toponium decays.

{Our results show that for the mass spectrum of toponium, the masses obtained using the Cornell potential are slightly higher than those calculated with the Coulomb potential, with the former also exhibiting marginally larger mass splittings. For the two-photon, two-gluon, and two-electron decay processes of the ground-state toponium, significant differences emerge between the results derived from the Cornell and Coulomb potentials. The decay partial widths obtained with the Coulomb potential are approximately twice those calculated using the Cornell potential, and our Coulomb potential results demonstrate good agreement with existing studies. The marked differences in theoretical results therefore indicate that future experimental investigations will readily discern whether the Coulomb or Cornell potential better describes the toponium system.}

{\bf Acknowledgments}
This work was supported in part by the National Natural Science Foundation of China (NSFC) under the Grants Nos. 12075073, 12075074, 12475077.

\begin{appendix}
{\section{Introduction of the BS equation and the Salpeter equation}\label{appendix}

The BS equation \cite{Salpeter:1951sz} for a meson with quark 1 and antiquark 2 is written as
\begin{equation}
\label{BS}
\chi_{_P}(q)
=iS(p_1)\int{\frac{d^4k}{(2\pi)^4}V(P,k,q)
\chi_{_P}(k)}S(-p_2),
\end{equation}
where $\chi_{_P}(q)$ is the relativistic wave function of the meson;
$S(p_1)$ and $S(-p_2)$ are the propagators of quark and antiquark; $V(P,k,q)$ is the interaction kernel.

In the instantaneous approximation, the kernel $V(P,q,k)$ becomes to $V(q_{_{P\bot}}-k_{_{P\bot}})$. We define the three dimensional wave function $\varphi(q_{_{P\perp}})$ and shorthand symbol $\eta_{_P}(q_{_{P\perp}})$,
\begin{equation}\varphi(q_{_{P\perp}}) \equiv i \int{\frac{dq_{_P}}{2\pi} \chi_{_P}(q)},
\qquad
\eta_{_P}(q_{_{P\perp}}) \equiv \int{\frac{dk^3_{_{P\perp}}}{(2\pi)^3} V(q_{_{P\bot}}-k_{_{P\bot}})\varphi(k_{_{P\perp}}) },\end{equation}
then the BS equation is changed to
\begin{equation}
\label{BS1}
\chi_{_P}(q)=S(p_1)\eta_{_P}(q_{_{P\perp}})S(-p_2),
\end{equation}
and the propagators can be written as
$$
iS(p_{1}) = \frac{\Lambda^+(p_{1_{P\perp}})}{p_{1_P}-\omega_{1}+i\epsilon} + \frac{\Lambda^-(p_{1_{P\perp}})}{p_{1_P}+\omega_{1}-i\epsilon},$$$$
-iS(-p_{2}) = \frac{\Lambda^+(-p_{2_{P\perp}})}{-p_{2_P}-\omega_{2}+i\epsilon} + \frac{\Lambda^-(-p_{2_{P\perp}})}{-p_{2_P}+\omega_{2}-i\epsilon},
$$
where the projection operators are
$$\Lambda^\pm(p_{1_{P\perp}})=\frac{1}{2\omega_{1}}\Big[\frac{\slashed P}{M}\omega_{1} \pm(m_{1}+\slashed p_{1_{P\perp}})\Big],
$$$$\Lambda^\pm(-p_{2_{P\perp}})=\frac{1}{2\omega_{2}}\Big[\frac{\slashed P}{M}\omega_{2} \pm(-m_{2}+\slashed p_{2_{P\perp}})\Big].$$

Using the contour integral method, after integrating over $q_{_P}$ on both sides of Eq.~\ref{BS1}, we obtain the Salpeter equation \cite{Salpeter:1952ib}
\begin{equation}\label{Salpeter}
\varphi(q_{_{P\perp}})=\frac{
\Lambda^{+}(p_{1_{P\perp}})\eta_{_P}(q_{_{P\perp}})\Lambda^{+}(-p_{2_{P\perp}})}
{(M-\omega_{1}-\omega_{2})}- \frac{
\Lambda^{-}(p_{1_{P\perp}})\eta_{_P}(q_{_{P\perp}})\Lambda^{-}(-p_{2_{P\perp}})}
{(M+\omega_{1}+\omega_{2})}\;.
\end{equation}
With the definitions
\begin{equation}\label{project}
\varphi^{\pm\pm}=
\Lambda^{\pm}(p_{1_{P\perp}})
\frac{\not\!{P}}{M}\varphi \frac{\not\!{P}}{M}
\Lambda^{{\pm}}(-p_{2_{P\perp}})\;,
\end{equation}
the wave function can be separated to four parts
\begin{equation}
\varphi(q_{_{P\perp}})=\varphi^{++}(q_{_{P\perp}})+
\varphi^{+-}(q_{_{P\perp}})+\varphi^{-+}(q_{_{P\perp}})
+\varphi^{--}(q_{_{P\perp}}),
\end{equation}
where $\varphi^{++}(q_{_{P\perp}})$ and $\varphi^{--}(q_{_{P\perp}})$ are called positive wave function and  negative wave function, respectively. The Salpeter equation Eq.~\ref{Salpeter} is also expressed as
\begin{equation}\label{posi}
\varphi^{++}(q_{_{P\perp}})=\frac{
\Lambda^{+}(p_{1_{P\perp}})\eta(q_{_{P\perp}})\Lambda^{+}(-p_{2_{P\perp}})}{(M-\omega_{1}-\omega_{2})}\;,
\end{equation}
\begin{equation}\label{nega}\varphi^{--}(q_{_{P\perp}})=-\frac{
\Lambda^{-}(p_{1_{P\perp}})\eta(q_{_{P\perp}})\Lambda^{-}(-p_{2_{P\perp}})}{(M+\omega_{1}+\omega_{2})}\;,
\end{equation}
\begin{equation}
\varphi^{+-}(q_{_{P\perp}})=\varphi^{-+}(q_{_{P\perp}})=0\;,
\label{const}
\end{equation}
so the complete Salpeter equation includes four independent equations.
To obtain a relativistic wave function for a meson, for example the one in Eq.~\ref{e_0-} or Eq.~\ref{e_1-}, we need solving the complete Salpeter equation. While for a non-relativistic wave function when $\varphi^{++}(q_{_{P\perp}}) \gg \varphi^{--}(q_{_{P\perp}})$, only the first equation  Eq.~\ref{posi} needs to be solved.}

\end{appendix}


\begin{thebibliography}{99}
\bibitem{pdg}
S.~Navas \textit{et al.} (Particle Data Group),
Phys. Rev. D \textbf{110} (2024) 030001.
{ \bibitem{Pancheri}G.~Pancheri, J.-P. Revol and C. Rubbia, Phys. Lett. B \textbf{277} (1992) 518.}
\bibitem{Sumino}
Y.~Sumino and H.~Yokoya,
JHEP \textbf{09} (2010), 034; JHEP \textbf{06} (2016), 037 (erratum).
\bibitem{Kawabata} S.~Kawabata and H.~Yokoya, Eur. Phys. J. C \textbf{77} (2017) 323.
{ \bibitem{WanliJu}W.-L. Ju, G. Wang, X. Wang, X. Xu, Y. Xu and L.-L. Yang, JHEP \textbf{06} (2020) 158.}
\bibitem{Fabio}F.~Maltoni, C.~Severi, S.~Tentori and E.~Vryonidou, JHEP \textbf{03} (2024) 099.
{\bibitem{Fadin}V.~S.~Fadin and V.~A.~Khoze, JETP. Lett. \textbf{46} (1987) 525.}
\bibitem{ma}K.~Hagiwara, K.~Ma and H.~Yokoya, JHEP \textbf{06} (2016) 048.
\bibitem{Kaoru}K.~Hagiwara, Y.~Sumino and H.~Yokoya, Phys. Lett. B \textbf{666} (2008) 71.
\bibitem{Kiyo}Y.~Kiyo, J.~H.~Kuhn, S.~Moch, M.~Steinhauser and P.~Uwer, Eur. Phys. J. C \textbf{60} (2009) 375.
\bibitem{penin}A.~A.~Penin and M.~Steinhauser, Phys. Lett. B \textbf{538} (2002) 335.
\bibitem{Kiyo2}Y.~Kiyo and Y.~Sumino, Phys. Rev. D \textbf{67} (2003) 071501.
{ \bibitem{Buchmuller}W. Buchmuller and S. H. H. Tye, Phys. Rev. D \textbf{24} (1981) 132.}
{ \bibitem{Moxhay}P. Moxhay and J. L. Rosner, Phys. Rev. D \textbf{31} (1985) 1762.}
{ \bibitem{Fabiano2}N.~Fabiano, A. Grau and G.~Pancheri, Phys. Rev. D \textbf{50} (1994) 3173.}
\bibitem{beneke}M.~Beneke, Y.~Kiyo and K.~Schuller, Nucl. Phys. B \textbf{714} (2005) 67.
{ \bibitem{Kuhn1}J. H. Kuhn and P. M. Zerwas, Phys. Rept. \textbf{167} (1988) 321.}
{ \bibitem{Kuhn2}J. H. Kuhn and E. Mirkes, Phys. Rev. D \textbf{48} (1993) 179.}
{ \bibitem{Fabiano1} N. Fabiano, A. Grau and G. Pancheri, Nuovo Cim. A \textbf{107} (1994) 2789.}
{ \bibitem{Fabiano}N.~Fabiano, Eur. Phys. J. C \textbf{26} (2003) 441.}
{ \bibitem{Cakir}O. Cakir, R. Ciftci, E. Recepoglu and S. Sultansoy, Acta Phys. Polon. B \textbf{35} (2004) 2103.}
{ \bibitem{Kats}Y. Kats and M. D. Schwartz, JHEP \textbf{04} (2010) 016.}
{ \bibitem{Bigi}I. I. Y. Bigi and H. Krasemann, Z. Phys. C \textbf{7} (1981) 127.}
{ \bibitem{Artymowicz}P. Artymowicz, Acta Phys. Polon. B \textbf{15} (1984) 505. }
{ \bibitem{Yndurain}F. J. Yndurain, Nucl. Phys. B Proc. Suppl.  \textbf{93} (2001) 196.}
\bibitem{atlas}G.~Aad, \textit{et al.} (ATLAS Collaboration), Nature \textbf{633} (2024) 8030, 542.
\bibitem{cms}A. Hayrapetyan, \textit{et al.} (CMS Collaboration), Rept. Prog. Phys. \textbf{87} (2024) 117801.
{ \bibitem{han}T. Han, M. Low and T. A. Wu, JHEP \textbf{07} (2024) 192.}
{ \bibitem{Dong}Z. Dong, D. Goncalves, K. Kong and A. Navarro, Phys. Rev. D \textbf{109} (2024) 115023.}
{ \bibitem{Casas}J. A. Aguilar-Saavedra and J. A. Casas, Phys. Rev. Lett.  \textbf{133} (2024) 111801.}
{ \bibitem{han2}K. Cheng, T. Han and M. Low, Phys. Rev. D \textbf{111} (2025) 033004.}
{ \bibitem{atlas2}M. Aaboud, \textit{et al.} (ATLAS Collaboration), Phys. Rev. D \textbf{98} (2018) 012003.}
{\bibitem{cms2}V. Khachatryan, \textit{et al.} (CMS Collaboration), Phys. Rev. D \textbf{95} (2017) 092001.}
\bibitem{ma2}B.~Fuks, K.~Hagiwara, K.~Ma and Y.-J. Zheng, Phys. Rev. D \textbf{104} (2021) 034023.
\bibitem{Aguilar}J.~A.~Aguilar-Saavedra, Phys. Rev. D \textbf{110} (2024) 054032.
{ \bibitem{Estrada}F. J. Llanes-Estrada, e-Print: 2411.19180.}
{ \bibitem{Francener}R. Francener, V. P. Goncalves and D. E. Martins, e-Print: 2502.03295.}
{\bibitem{2411}L. Jeppe, (CMS Collaboration), e-Print: 2411.18414; Contribution to: TOP2024.}
\bibitem{Salpeter:1951sz}
E.~E.~Salpeter and H.~A.~Bethe,
Phys. Rev. \textbf{84} (1951), 1232.
\bibitem{Salpeter:1952ib}
E.~E.~Salpeter,
Phys. Rev. \textbf{87} (1952), 328.
\bibitem{Kim:2003ny}C.~S.~Kim and G.-L. Wang, Phys. Lett. B \textbf{584} (2004) 285; Phys. Lett. B \textbf{634} (2006) 564 (erratum).
\bibitem{Wang:2005qx}G.-L. Wang, Phys. Lett. B \textbf{633} (2006) 492.
{\bibitem{cornell}E. Eichten, K. Gottfried, T. Kinoshita, K. D. Lane and T. M. Yan, Phys. Rev. D \textbf{21} (1980) 203.}
{\bibitem{Richardson}J.~L.~Richardson, Phys. Lett. B \textbf{82} (1979) 272.}
{\bibitem{ktchao}J. Tang, J.-H. Liu and K.-T. Chao, Phys. Rev. D \textbf{51} (1995) 3501.}
{ \bibitem{Goncharov}Y. P. Goncharov, Nucl. Phys. A \textbf{808} (2008) 73.}
\bibitem{Wang:2022cxy}G.-L. Wang, T. Wang, Q. Li and C.-H. Chang, JHEP \textbf{05} (2022) 006.
{\bibitem{wglP}G.-L. Wang, Phys. Lett. B \textbf{650} (2007) 15.}
{\bibitem{wglP2}G.-L. Wang, Phys. Lett. B \textbf{674} (2009) 172.}
{\bibitem{tianh}T. Wang, G.-L. Wang, W.-L. Ju and Y. Jiang, JHEP \textbf{03} (2013) 110.}
{\bibitem{tianh2} T. Wang, H.-F. Fu, Y. Jiang, Q. Li and G.-L. Wang, Int. J. Mod. Phys. A \textbf{32} (2017) 06\&07, 1750035.}
{ \bibitem{Enterria}D. d'Enterria and K. Kang, e-Print: 2503.10952.}
\end{thebibliography}

 \end{document}